\documentclass[fleqn,usenatbib,useAMS]{mnras}

\usepackage{graphicx}  % figures
\usepackage[english]{babel}
\usepackage{amsmath,amssymb}  % math symbols
\usepackage{xspace}  % to avoid spaces being eaten by custom commands
\usepackage{xcolor}
\usepackage{threeparttable}  % for table notes

\usepackage[T1]{fontenc}
\usepackage{ae,aecompl}
\usepackage{newtxtext,newtxmath}

\graphicspath{ {figs/} } 

\newcommand{\J}{\mbox{J0533$-$4524}\xspace}  % pulsar name

\newcommand{\pccm}{{\ensuremath\mathrm{\,pc\,cm^{-3}}}}  % pc /cc
\newcommand{\Msun}{\ensuremath{\mathrm{M}_\odot}}  % solar mass

\DeclareMathOperator\erfc{erfc} % error function
\DeclareMathOperator\erfcinv{erfc^{-1}}  % inverse error function

\title[sdB-PSR binaries and discovery of PSR~\J]{A search for pulsars in subdwarf B binary systems \\
  and discovery of giant-pulse emitting PSR~\J}

\author[L.C. Oostrum et al.]{
L. C. Oostrum$^{1,2}$\thanks{E-mail: l.c.oostrum@uva.nl},
J. van Leeuwen$^{1,2}$,
Y. Maan$^{1}$,
T. Coenen$^{2,1}$,
C.H. Ishwara-Chandra$^{3}$
\\
$^{1}$ASTRON, The Netherlands Institute for Radio Astronomy, Oude Hoogeveensedijk 4, 7991 PD, Dwingeloo, The Netherlands\\
$^{2}$Anton Pannekoek Institute for Astronomy, University of Amsterdam, PO Box 94249, 1090 GE, Amsterdam, The Netherlands\\
$^{3}$National Centre for Radio Astrophysics, TIFR, Post Bag 3, Pune University Campus, 411007 Pune, India
}

\date{Accepted 2020 January 13. Received 2020 January 13; in original form 2019 October 11}

\pubyear{2020}

\begin{document}
\maketitle

\begin{abstract}
Binary millisecond pulsars (MSPs) provide several opportunities for research of fundamental physics. However, finding them can be challenging. Several subdwarf B (sdB) binary systems with possible neutron star companions have been identified, allowing us to perform a targeted search for MSPs within these systems.
Six sdBs with companions in the neutron star mass range, as determined from their optical light curves, were observed with the Green Bank and Westerbork radio telescopes. The data were searched for periodic signals as well as single pulses.
No radio pulsations from sdB systems were detected, down to an average sensitivity limit of $0.11\,$mJy. 
We did, however, discover a pulsar in the field of sdB HE0532$-$4503. Follow-up observations with the Giant Metrewave Radio Telescope showed that this pulsar, \J,
is not spatially coincident with the sdB system.
The pulsar has a relatively low magnetic field but still emits giant pulses.
We place an upper limit of three to the number of radio pulsars in the six sdB systems. The non-detections  may be explained by a combination of the MSP beaming fraction, luminosity, and a recycling fraction <0.5. Alternatively, the assumption of co-rotation between the MSP and sdB may break down, which implies the systems are more edge-on than previously thought. This would shift the predicted companion masses into the white dwarf range. It would also explain the relative lack of edge-on sdB systems with massive companions.
\end{abstract}


\maketitle

\section{Introduction}
In binary systems, accretion may convert normal pulsars (PSRs) into fast-spinning, low magnetic field, millisecond pulsars (MSPs). 
Timing pulse arrivals from pulsars in such systems in order to extract their properties 
offers tests and insights in a number of fundamental physics areas. 
One can constrain neutron star masses and equations of state \citep{lp01}, study binary evolution, and strong field general relativity
(if the pulsar is in orbit with a massive companion, e.g. \citealt{tw89}). Furthermore, gravitational radiation from distant supermassive black hole mergers can potentially be detected
using pulsar timing arrays made up of stably rotating and emitting pulsars \citep{jb03}.

Finding new MSPs in a blind, wide-field survey is a challenge.
Blind surveys for radio pulsars have lead to discoveries of numerous MSPs
\citep[e.g.][]{2015MNRAS.446.4019B,2019A&A...626A.104S,2019ApJ...886..148P}.
Targeted searches allow for an increased sensitivity and a more efficient use of telescope time \citep[see, for example,][]{2015ApJ...810...85C,2018ApJ...864...16M}.
As MSPs need a binary companion for their formation, selecting targets based on potential companions identified at optical wavelengths promises to be efficient. Generally these companions are assumed to be 
white dwarfs \citep{lfm+07,2009ApJ...697..283A}, but MSP - subdwarf systems are also possible. In those systems, a sub-luminous B dwarf star (sdB) would have spun up the MSP.
% The unique formation mechanism of sdB stars potentially creates remarkable binary systems where the orbit is tight, even
% relativistic. This suggests MSPs found in such targeted binaries may help constrain theories of gravity
% \citep[cf.][]{2012MNRAS.423.3328F}.

Sub-luminous B dwarfs (sdBs) are thought to be light (${\sim} 0.5\Msun$), core helium-burning stars. In contrast to main-sequence core helium-burning stars, they have thin hydrogen envelopes \citep{heber1986} and a peculiar composition that does not fit the usual MK classification scheme \citep{drilling2013}.
A large fraction of sdBs -- up to 2/3 for some surveys \citep{2001MNRAS.326.1391M} -- are in tight (hours to days) binaries, 
but not all are. Thus, while binary evolution must be important in the development of these stars, their exact
  formation is a matter of ongoing debate. The detection of an MSP companion would therefore not only advance research into
  compact objects; it could also help explain the formation of subdwarf stars.
Among the ${\sim}6000$ hot subdwarf stars known in our Galaxy, ${\sim}3500$ are sdB stars \citep{2017A&A...600A..50G}.
There may be several million sdB stars, although a significant fraction may have been missed so far due to selection effects \citep{hpm+03}.
Several channels of binary evolution could have been responsible for producing the observed sdB population:
Stable Roche-Lobe overflow could lead to a longer-period binary with a main-sequence companion \citep{hpm+03,chd+13}. They might also be formed through the merger of two white dwarfs, resulting in an isolated sdB \citep{web84}.
Lastly, a common envelope channel could lead to an sdB with a massive compact companion \citep{geier+2010}.
This channel involves a wide binary with a massive primary, that will go on to
form a neutron star (NS) or black hole (BH), two phases of common envelope evolution and a short X-ray binary phase.
A first
common envelope phase starts soon after the primary reaches the red supergiant stage of its evolution and starts overflowing its Roche Lobe.
During this first common envelope phase the binary tightens. The second phase of mass transfer starts when the secondary begins to overflow its
Roche Lobe. If the primary, which by then has undergone a supernova, is a neutron star it will get recycled. 
The second common-envelope phase
starts shortly thereafter, tightens the binary further and dissipates the envelope of the secondary. The secondary, which is now mostly
stripped of its hydrogen envelope, continues its evolution as an sdB star.
For a recent review on sdB stars see \citet{heber2016}.

  Based on population synthesis,
  \citet{2005ARep...49..871Y} find that in 9 out of 10 observed sdB binaries,
  the companion is a white dwarf. In 1 out
  of 10 it is a main sequence star. For  compact object companions, simulations by
  \citet{2010Ap&SS.329...25N} show that 1 in 100 sdB binaries contains a neutron
  star, and 1 in 10,000 a black hole.
  Thus, of the known sdB stars, a few dozen are expected to orbit a neutron star;
  a handful may be in a binary with a black hole.   
  So far, no such compact-object companion has been directly confirmed.
Positively identifying a pulsar in a tight, hours to days, sdB binary would provide constraints on binary evolution leading to
these systems \citep[see e.g.][]{cls11}.
The unique formation mechanism of sdB stars potentially creates remarkable binary systems where the orbit may even be relativistic. This suggests MSPs found in such targeted binaries may help constrain theories of gravity \citep[cf.][]{2012MNRAS.423.3328F}.
Furthermore, the timing of such an MSP would inform us of the size of the projected neutron-star orbit and the neutron-star
  velocity. Combined with a canonical neutron-star mass and measurement of the sdB velocity, this would provide a derivation of the sdB mass.
If the MSP timing stability allows for the determination of  post-Keplerian parameters,
  as in e.g., \citet{2015ApJ...798..118V},
  one can deduce these parameters more precisely still, to about 1\%. 

Given sufficiently deep observations, non-detections
of radio pulsations from these systems could mean the absence of a neutron star,
but may also be explained by a pulsar that is either off (because, for example, it was insufficiently recycled),
or beamed away from Earth. Non-detections in a large enough sample of sdB stars provide statistics on the sdB formation channels. 

In Sect.~\ref{sec:cands} our target selection is described. Section~\ref{sec:obs} gives an overview of our observations and data reduction. We show the results on sdB systems and the discovery of a new pulsar in Sects.~\ref{sec:resultssdb} and \ref{sec:resultshe0532}. In Sect.~\ref{sec:discussion} we discuss our findings and in Sect.~\ref{sec:conclusions} we show our conclusions.

\section{Candidate selection}\label{sec:cands}
Our targets were selected from a sample of sdB stars presented in \citet{geier+2010}. Using multiple optical spectra spread over the orbit of the
sdBs, they determine the radial velocity curves and hence constrain the mass function:
\begin{equation}
    f_\mathrm{m} = \frac{M_\mathrm{comp}^3\sin^3i}{\left(M_\mathrm{comp}+M_\mathrm{sdB}\right)^2} = \frac{P K^3}{2\pi G}\,,
\end{equation}
where $M_\mathrm{sdB}$ and $M_\mathrm{comp}$ are the masses of the sdB and its companion, $i$ is the inclination angle of the orbital plane, $P$
is the orbital period, and $K$ is the radial velocity semi-amplitude of the sdB.

Under the assumption that the sdBs are tidally locked, the inclination can be determined from their observed rotational velocities.
Furthermore, in a binary, the sdB mass resulting from the common-envelope ejection channel is predicted to be in a very narrow range of 0.46 - 0.50 $\,\mathrm{\Msun}$, with a canonical value of $0.46 \,\mathrm{\Msun}$ \citep{hpm+02,hpm+03}. However, in those cases the companion is typically a white dwarf. For neutron star companions, the allowed mass range may be larger, 0.3-1.1 $\,\mathrm{\Msun}$ \citep{geier+2010}. For the systems discussed in this paper, \citet{geier+2010} could not determine the sdB mass but used a canonical value of $0.5\,\mathrm{\Msun}$.
The derived inclinations then led to predictions for the companion masses \citep{geier+2010}.

Out of the 31 systems for which an estimate for the companion mass was determined, six have companions whose
  derived mass is above the Chandrasekhar limit, hence these are considered candidate neutron stars. Four of these are
at least 1$\sigma$ above the Chandrasekhar limit. The common envelope channel predicted to lead to such sdB binary
  systems with a massive compact companion also suggests mass transfer onto the compact object \citep{geier+2010}. As
  this is the canonical scenario for creating an MSP \citep{rs82,acrs82}, the neutron star candidates are considered to
  be MSP candidates as well. As the masses are the only argument for the companions being neutron stars, some of them might, however, be white dwarfs (WDs). 

Three of the candidates, HE\,0929$-$0424, HE\,0532$-$4503, and PG\,1232$-$136, were already observed and analysed in \citet{cls11}. No pulsations
were found there, and the pseudo-luminosity of any recycled pulsar in HE\,0929$-$0424 and PG\,1232$-$136 was strongly constrained. A weaker constrain was put on HE\,0532$-$4503. Here we present an analysis of
deeper observations of the same sources, as well as of three additional sources: PG\,1101+249, PG\,1432+159, and PG\,1743+477. While PG\,1232$-$136 is expected to host a black hole as its derived companion mass is a lower limit of $6\Msun$ \citep{geier+2010}, it might host a massive neutron star if the sdB in the binary system is not tidally locked.
An overview of the targets is given in Table~\ref{tab:targets}.

\begin{table*}
\centering
\caption{Overview of target parameters (based on \citealt{geier+2010}, distances from Gaia~Data~Release~2; \citealt{gaiadr2}) and observations. GBT observations were carried out with GUPPI at 300-400\,MHz, for WSRT we used PUMA{\sc ii} at 310-375\,MHz. $S_\mathrm{min}$ is the minimum detectable flux density for the full observation duration.}
\label{tab:targets}
\renewcommand{\arraystretch}{1.4}  % increase vertical separation to accommodate upper and lower errors
\setlength{\tabcolsep}{5pt}% decrease horizontal separation to make the table fit on the page
\begin{tabular}{*{12}l}
\hline
Target & $l$ & $b$ & $P_\mathrm{orb}$ & $i$ & $M_\mathrm{comp}$ & Companion & Distance & Telescope & MJD & $t_\mathrm{obs}$ & $S_\mathrm{min}$ \\
 & ($\degr$) & ($\degr$) & (d) & ($\degr$) & ($M_\odot$) & & ($\mathrm{kpc}$) & & & (hr) & ($\mathrm{mJy}$)\\ \hline
HE\,0929$-$0424     & 238.52 & +32.36 & 0.44 & $23^{+5}_{-4}$ & $1.82^{+0.88}_{-0.64}$ & WD/NS/BH & 1.7(3) & GBT & 55687.03, 55690.91$^\mathrm{(a)}$ & 0.5 + 0.8 & 0.11$^\mathrm{(b)}$\\
HE\,0532$-$4503     & 251.02 & $-$32.14 & 0.27 & $14^{+2}_{-2}$ & $3.00^{+0.94}_{-0.92}$ & NS/BH & 2.9(5) & GBT & 55852.34 &  2.1 & 0.07\\
PG\,1101+249        & 212.76 & +65.88 & 0.35 & $26^{+6}_{-4}$ & $1.67^{+0.77}_{-0.58}$ & WD/NS/BH & 0.43(1) & GBT & 55715.00 & 2.0 & 0.07\\
PG\,1232$-$136      & 296.99 & +48.77 & 0.36 & ${<}14$ & ${>}6.00$ & BH & 0.50(1) & GBT & 55687.08 & 0.9 & 0.12\\
PG\,1432+159        & 012.83 & +63.32 & 0.22 & $16^{+5}_{-3}$ & $2.59^{+2.01}_{-1.10}$ & NS/BH & 0.63(3) & WSRT & 55634.89 & 3.8 & 0.22\\
PG\,1743+477        & 074.41 & +30.66 & 0.52 & ${<}27$ & $>1.66$ & NS/BH & 0.77(2) & WSRT & 55650.27 & 3.8 & 0.21\\ \hline
\end{tabular}
\begin{tablenotes}
\item $^\mathrm{(a)}$ This source was observed in two separate sessions.
\item $^\mathrm{(b)}$ For the longest observation of this source.
\end{tablenotes}
\end{table*}

\section{Observations and data reduction}\label{sec:obs}
The six targets were observed with either the Westerbork Synthesis Radio Telescope (WSRT) at 310-375\,MHz,
or the Robert C. Byrd Green Bank Telescope (GBT) at 300-400\,MHz
as shown in Table~\ref{tab:targets}.
WSRT is a tied array of 14 25-m dishes spread over a 3-km baseline,
  resulting in a beam size of $1\arcmin$ at 350\,MHz.
  At the same frequency, the 100-m GBT has a beam size of $35\arcmin$.
For each target, the observation duration was chosen such that 
  we would have detected 95-100\% of the known MSPs when placed at the source distance 
  \citep[i.e., close to full completeness, see][]{cls11}.
In each observing session, a bright pulsar was observed as a test source to ensure that our 
observing setup and data reduction pipeline performed as expected.

The follow-up observations of a strong candidate from our search were performed using the GBT as well as the upgraded Giant Metrewave Radio Telescope \citep[uGMRT;][]{gak+17}, and these are discussed in more detail in Sect.~\ref{sec:resultshe0532}.

All data were searched with the {\sc PRESTO}\footnote{\url{https://github.com/scottransom/presto}} package \citep{presto}.
We used {\sc rfifind} to create radio frequency interference (RFI) masks which were used with subsequent processing. For uGMRT data, any strong \textsl{periodic} RFI (such as 50\,Hz interference from the power lines) were identified and excised from the individual frequency channels using {\sc rfiClean}\footnote{\url{https://github.com/ymaan4/rfiClean}} (Maan \& van Leeuwen in prep.).

Using the NE2001 \citep{cl02} and YMW16 \citep{ymw16} Galactic electron density models,
we converted the sdB distances to an expected dispersion measure (DM) towards each source. Based on this, we chose a DM upper limit in our search of $500\pccm$, which is well above the expected value of
${\sim}50\pccm$ predicted by the models. A higher value was chosen to account for the factor few uncertainty in Galactic electron density models,
as well as uncertainties in the distance to the sources.
Using {\sc DDplan.py} from {\sc PRESTO},
a dedispersion plan was determined for each observation,
taking into account the different time resolution and channel width for each
instrument
and optimised for minimal computing time at maximum resolution and sensitivity.
This led to typically 10000 timeseries over a DM range of 0-500$\pccm$, with DM steps of $0.01\pccm$ for DMs ${<}35\pccm$
up to $0.3\pccm$ for DMs ${>}400\pccm$.
Each resulting timeseries was searched for both single pulses and periodic signals, as described hereafter.

\subsection{Periodicity search}
The periodicity search was done in the frequency domain, using {\sc accelsearch} from {\sc PRESTO}.We searched for spin periods between $0.1\,$ms and $1\,$s.
Up to 16 harmonics were summed to improve sensitivity to narrow
pulses. The likely strong acceleration of the targets in their binary orbits causes a drift of the signal in Fourier
space.
This drift can be searched for by {\sc accelsearch}, but only in the regime of constant acceleration.
That assumption is typically valid if the observation is shorter than 10\% of the orbital period.
Most of our observations are, however, longer.
For those, we searched both the full data, as well as chunks of at most 10\% of the orbit.

The maximum expected line-of-sight acceleration ($a_\mathrm{max}$) is given by
\begin{equation}
\label{eq:accelmax}
a_\mathrm{max} = \Omega_\mathrm{b} r_\mathrm{p} = \left[\frac{\mathrm{G} M_\mathrm{sdB}^3 \Omega_\mathrm{b}^4}{\left(M_\mathrm{sdB}+M_\mathrm{p}\right)^2}\right]^{1/3},
\end{equation}
where $\Omega_\mathrm{b}$ and $r_\mathrm{p}$ are the mean angular velocity and semi-major axis of the pulsar orbit, $M_\mathrm{sdB}$ and
$M_\mathrm{p}$ are the masses of the sdB and pulsar, and G is the gravitational constant. The second equality is given by Kepler's 3rd law, which
is valid given the observed non-relativistic orbital velocities of the sdBs. The suspected neutron star companions are all more massive than the
sdBs and hence have a lower orbital velocity. For a canonical pulsar of mass $1.4\Msun$ and sdB of mass $0.46\Msun$, the maximum acceleration
ranges from $11\,\mathrm{m\,s^{-2}}$ for PG\,1743+477 to $35\,\mathrm{m\,s^{-2}}$ for PG 1432+159. We use these values to set the maximum Fourier space drift in {\sc accelsearch}, which then searches for accelerations between zero and the given value.

The candidates produced by {\sc accelsearch} are sifted using {\sc ACCEL\_sift.py} from {\sc PRESTO} and each candidate with a signal-to-noise ratio ($S/N$) of 8 was folded on the raw data and
visually inspected.

\subsection{Single pulse search}
Each timeseries was searched for single pulses using {\sc single\_pulse\_search.py} from {\sc PRESTO}. The $S/N$ at each point in the timeseries is determined using a Fourier-domain matched-filter technique, with boxcar widths between one sample and the number of samples corresponding to $20\,\mathrm{ms}$. This means the search is sensitive to single pulses with widths up to $20\,\mathrm{ms}$. The matched-filtering is not sensitive to the phase of the pulse. All single-pulse candidates above a $S/N$ of 8 were visually inspected.

\section{Results}\label{sec:resultssdb}
All test pulsars were successfully detected by our pipeline. The single pulse search yielded similar results. The only test pulsar known to emit giant pulses, PSR B1937+21 \citep{cst+96}, was
blindly re-detected.
In our sample of six sdB systems, one pulsar candidate is identified towards sdB
HE0532$-$4503 with a period of $157.28\,$ms and DM of $19\pccm$ (Fig.~\ref{fig:J0533_discovery}). In addition, three single pulses were detected towards HE\,0532$-$4503 (Fig.~\ref{fig:j0533_gp}). These were detected at the same DM as the periodic candidate.

% We hypothesised the visible sdB star spun up the pulsar,
% since the sdB star  must be the secondary and did not explode in a supernova.
% The period was longer than expected for a fully recycled MSP, however, at $157.28\,$ms.
% Additionally, there was no measurable  orbital acceleration. We sought to confirm the pulsar and start a timing solution to provide more insight into these concerns.
The system was observed several times with GBT, Parkes and uGMRT (Table~\ref{tab:he0532_followup}). Periodic emission from the pulsar was not detected in the first three follow-up observations on 20160528 to 20160530. Single pulses continued to be visible.
From timing on these single pulses, and later on periodic detections,
we learned the pulsar was isolated and was found by chance in the sdB star field.
The discovery of this pulsar, PSR~\J, is further discussed in Sect.~\ref{sec:resultshe0532}.

We thus detect no pulsars in any of the observed sdB binary systems. This does not directly imply none of these sdB
systems host a pulsar. They could be too faint, or their emission could be beamed away from Earth.

An upper limit to the flux density of any pulsar beamed towards Earth can be set using the radiometer equation \citep{dtws85,lk05}:
\begin{equation}
S_\mathrm{min} = \frac{(S/N)_\mathrm{min} \, T_\mathrm{sys}}{G \sqrt{N_\mathrm{pol} \, BW \, T_\mathrm{obs}}} \, \sqrt{\frac{W}{P-W}},
\end{equation}
Where $S_\mathrm{min}$ is the minimum detectable flux density, $(S/N)_\mathrm{min}$ is the $S/N$ threshold in the search, $T_\mathrm{sys}$ is the
sum of the receiver temperature ($T_\mathrm{rec}$) and the sky temperature ($T_\mathrm{sky}$), G is the telescope gain, $N_\mathrm{pol}$ is the
number of polarisations (=2), $BW$ is the bandwidth, $T_\mathrm{obs}$ is the length of the observation, $P$ is the pulsar period and $W$ is width of
the pulse profile. As the putative pulsars are expected to be MSPs (here defined, following \citealt{m17}, as pulsars
with $P {<} 100\,$ms and with $\dot{P} {<} 10^{-17}$), we adopt the median value of $W/P = 0.08$ of MSPs in the ATNF
pulsar catalogue \citep{hm04}, where width is defined as the width at 50\% of the maximum of the pulse profile.
While this duty cycle is somewhat larger at lower frequencies -- suggesting the  beam illuminates a larger
  part of the celestial sphere over the pulsar, and increasing the odds of Earth being in it -- the difference is
  negligible compared to the other uncertainties in determining $S_\mathrm{min}$.
The sky temperature is taken from the \citet{hssw82} sky map, scaled from $400\,\mathrm{MHz}$ to the central frequency of each instrument using a scaling of
 $T_\mathrm{sky} \propto \nu^{-2.6}$ \citep{lmo+87}. For GBT we use $T_\mathrm{rec}$ = $58\,$K and $G$ = $2.0\,$K/Jy \footnote{GBT observer guide
  \url{https://science.nrao.edu/facilities/gbt/observing/GBTog.pdf}}, for WSRT $T_\mathrm{rec}$ = $125\,$K  and
$G$ = $1.1\,$K/Jy 
\citep{2013MNRAS.428.2857R}.
the obtained flux density limits are listed in Table~\ref{tab:targets}.

We run a Monte-Carlo (MC) simulation to determine how many pulsars we expect to detect. We use a pseudo-luminosity distribution following a log-normal distribution with mean $-1.1$ and standard deviation $-0.9$ \citep{fk06}. This distribution was determined for normal pulsars at $1400\,$MHz, but later shown to be valid for recycled pulsars in globular clusters \citep{blc11}. The distribution is scaled from 1400\,MHz to 350\,MHz using a spectral index of $-1.9$, which is a typical value as used in \citet{blc11}.

The luminosity distribution gives the probability that a pulsar is bright enough to be detected from Earth,
but ignores any beaming effect.
The beams of MSPs are larger than those of normal pulsars,
and their beaming fractions $f_b$ are typically as high as 0.5 to 0.9 \citep{kxk+98}.
Still these do not cover the entire sky so there is a 10-50\% chance the beam misses Earth. In our simulation, we assume a uniform distribution of $f_b$ between 0.5 and 0.9.

For the MC simulation, we first assign $N$ out of the six systems to actually host a pulsar. For each value of $N$ from zero to six, we simulate $N \times 25000$ pulsars. Each pulsar is randomly assigned to one of the sdB systems, from which then a distance is drawn using a normal distribution with mean and standard deviation equal to the values determined for the sdB by Gaia (\citealt{gaiadr2}, see also Table~\ref{tab:targets}). The pulsar is also assigned a pseudo-luminosity and beaming fraction following the above described distributions. The pulsar is considered detected if (i) the flux density determined from the pseudo-luminosity and sdB distance is above the threshold for the corresponding observation (see Table~\ref{tab:targets}) and (ii) the pulsar beam sweeps across Earth, which is true with a probability equal to the beaming fraction.

When using the flux density threshold as determined for the full observation duration (cf. Table~\ref{tab:targets}), we find that if more than three systems host a pulsar, we would have detected at least one in $>$97\% of the iterations. As we also searched the data in chunks of at most 10\% of the orbital duration of the sdBs to avoid strong acceleration effects, we repeat the simulation with flux density thresholds determined from the duration of those chunks. Then, we would have detected at least one pulsar in $>$95\% of the iterations. We thus conclude that \emph{at most} three of the sdB systems host an MSP. If they do host MSPs, these must either be very faint or their beam does not sweep across the Earth.

If pulsars in sdB systems are only mildly recycled, their beams may be larger. The beaming fraction is also the most important factor in the number of detectable pulsars; for a uniform beaming fraction distribution of 0.3-0.5, it is possible that all six systems host a pulsar at the 95\% confidence level.

\section{Discovery of PSR J0533$-$4524}\label{sec:resultshe0532}
\begin{figure}
    \centering
    \includegraphics[width=\columnwidth]{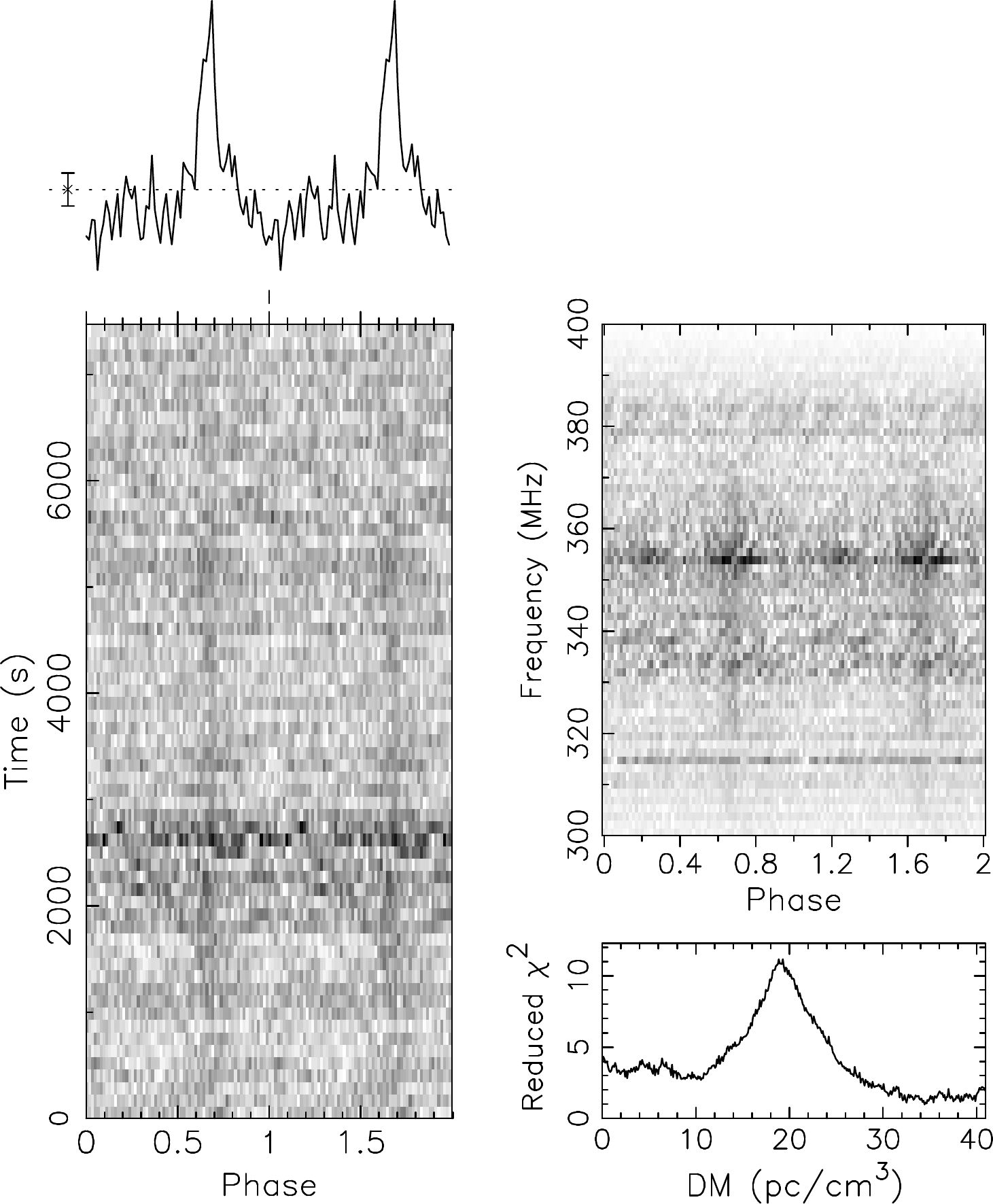}
    \caption{Discovery of PSR \J with GBT. The signal is strongly peaked (top-left, two rotational periods
      shown),  detected in the first 1.7 hrs out of the 2.1-hr observation
       (bottom-left) and broadband (top-right), with a well-defined
      dispersion measure (bottom-right).}
    \label{fig:J0533_discovery}
\end{figure}

\begin{figure}
    \centering
    \includegraphics[width=\columnwidth]{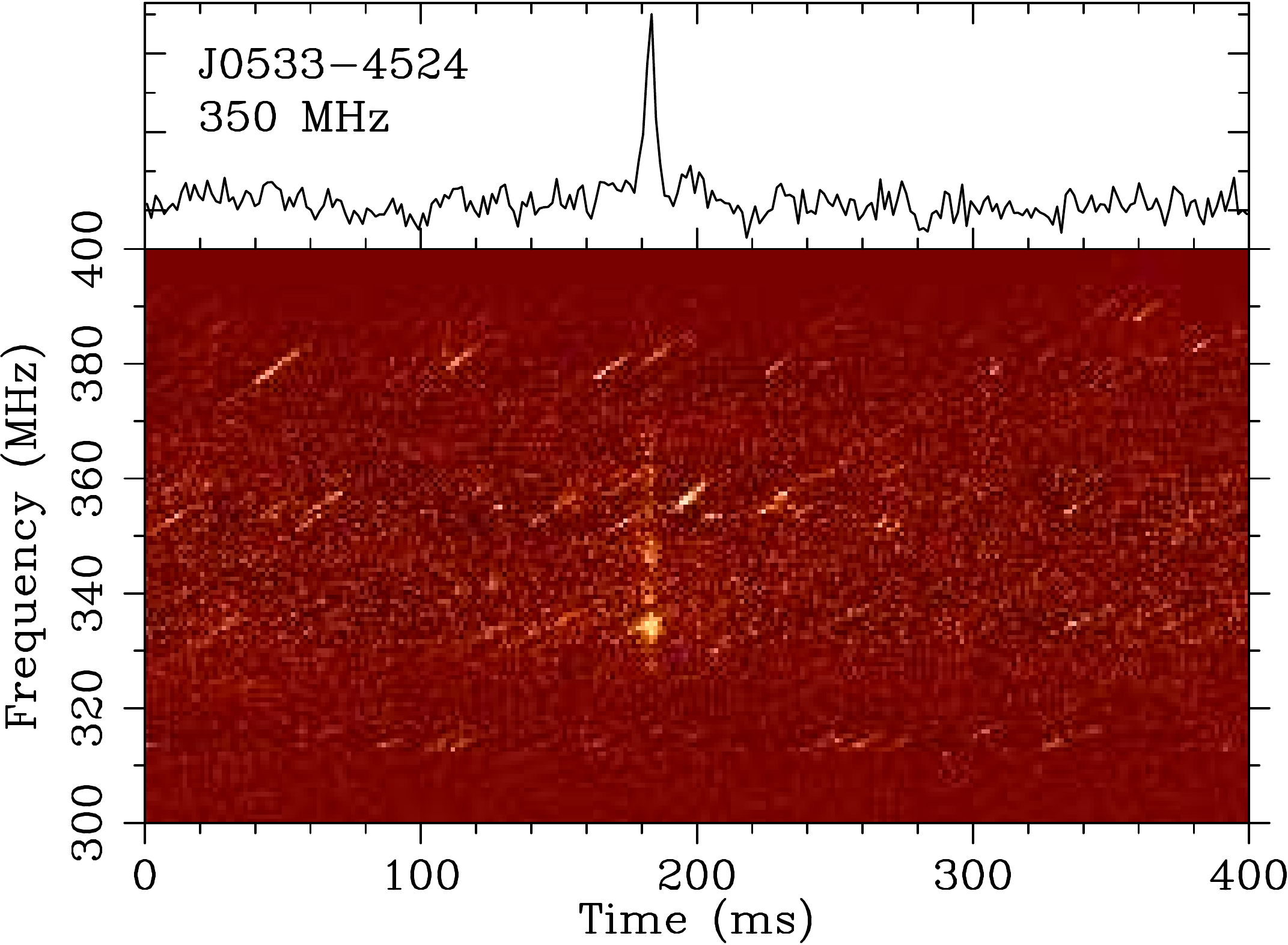}
    \caption{Brightest PSR \J giant pulse detected with GBT, with a peak flux density of $3.7\,$Jy and fluence of $19\,$Jy-ms.}
    \label{fig:j0533_gp}
\end{figure}

One convincing pulsar candidate was detected in our data, towards sdB HE0532$-$4503. The candidate was detected with a $S/N$ of $22$ at a period of $157.28\,$ms and a DM of $19\pccm$. The periodic pulse profile is shown in Fig.~\ref{fig:J0533_discovery}. The signal is broadband and clearly visible throughout most of the observation.
In addition to the periodic signal, three single pulses were detected with a $S/N$ between 10 and 30, all of which reached a maximum $S/N$ at the DM of the periodic candidate. The brightest detected pulse is shown in Fig.~\ref{fig:j0533_gp}.

The visible sdB star must be the secondary in the HE0532$-$4503 binary system,
and since it did not explode in a supernova, we assumed it spun up the pulsar.
As we were expecting to find an MSP \citep[see, e.g.,][]{2018A&A...618A..14W},
the period of the newly-found pulsar was somewhat long, at $157.28\,$ms.
Perhaps the
second stage of mass transfer was interrupted relatively quickly for the common-envelope stage?
The sdB-PSR association hypothesis was
furthermore challenged by the absence of measurable acceleration in the initial 2.1-hr observation,
a significant fraction of the 6.5-hr orbit. Perhaps the sdB star was lighter than expected?
We aimed to quickly confirm the pulsar and localise it through timing to answer these questions.

Under director's discretionary time, we observed the system five more times with GBT and we obtained several observations with uGMRT. An overview of the follow-up observations is given in
Table~\ref{tab:he0532_followup}. The table also shows the $S/N$ of the detected periodic signal when detected, as well as the number of detected single pulses. 

The periodic signal was detected in three out of six GBT observations, confirming that the candidate is indeed a real pulsar. We next aimed to localise the pulsar to
determine whether or not it could be part of the sdB binary system.

\begin{table*}
\centering
\caption{Overview of observations of PSR \J. For completeness, we list not only the follow-up observations but also the original discovery observation taken on 20111018.
For GBT observations, we used GUPPI at 300-400\,MHz. With uGMRT, we used GWB at 300-500\,MHz, and we recorded
  coherently and incoherently beamformed data, as well as interferometric data. The periodic flux density was determined
  using the radiometer equation, with the sensitivity scaled to the position of the pulsar (see
  Sect.~\ref{sec:sub:localisation}) in the beam. We assume 20\% errors on these flux densities. The number of single
  pulses above $S/N=8$ are listed in the 6\textsuperscript{th} column. 
  Finally, we indicate the orbital phase of HE0532$-$4503, assuming an orbit of $0.2656 \pm 0.0001$\,d
  \citep{geier+2010}, at the mid-point of the observation, relative to the 20160528 GBT observation. We only list these
  for epochs spaced closely enough in time to have an error on the orbital phase of less than 0.05.
  In addition to the observations listed here, we obtained one observation with Parkes, but as described in
  Sect.~\ref{sec:sub:localisation}, it had no significant sensitivity in the direction of the pulsar so we did not consider it further.
}
\label{tab:he0532_followup}
\begin{tabular}{*{10}l}
\hline
Date & MJD & Telescope & t$_\mathrm{obs}$ & $S/N$ & Periodic average flux density & Number of single pulses & Relative orbital phase\\ 
     &      &          &   (hr)   &    &  (mJy) & & (0.0-1.0) \\ \hline
20111018 & 55852.3417 & GBT    & 2.1 & 22 &  1.02(20) & 3 & \\
%20160405 & Parkes & 6.7 & $<$6 & <45(9)$^\mathrm{(a)}$ & 0 & \\
20160528 & 57536.76 & GBT    & 1.6 & $<$6 & <0.32(6) & 0 & 0.0 \\
20160529 & 57537.75 & GBT    & 1.5 & $<$6 & <0.32(6) & 2 & 0.76 \\
20160530 & 57538.75 & GBT    & 1.5 & $<$6 & <0.33(7) & 4 & 0.50\\
20160531 & 57539.75 & GBT    & 1.5 & 7 & 0.37(7) & 4 & 0.26 \\
20160805 & 57605.56 & GBT    & 1.6 & 11 & 0.55(11) & 5 & \\
20161106 & 57697.80 & uGMRT  & 4.5 & 19, 7$^\mathrm{(a)}$ & 0.44(9) & 11 & \\
20170108 & 57761.65 & uGMRT  & 3.9 & $<$6 & <0.16(3) & 0 & \\
20170302 & 57814.57 & uGMRT  & 2.2 & $<$6 & <0.20(4) & 0 & \\ 
20181103 & 58424.93 & uGMRT  & 1.6 & $<$6 & <0.20(4) & 0 & \\ 
20181201 & 58452.78 & uGMRT  & 1.9 & $<$6 & <0.18(4) & 0 & \\ 
20190103 & 58486.66 & uGMRT  & 1.1 & $<$6 & <0.24(5) & 0 & \\ 
20190201 & 58515.67 & uGMRT  & 1.7 & 50$^\mathrm{(b)}$ & 0.58(10) & 120 & \\ 
20190301 & 58543.50 & uGMRT  & 1.7 & $<$6 & <0.05(1) & 6 & \\
\hline

\end{tabular}
\begin{tablenotes}
%\item $^\mathrm{(a)}$ At $1.4\,$GHz, using $G = 0.83\,$K/Jy and $T_\mathrm{rec}=25\,$K from the Parkes user guide (\url{https://www.parkes.atnf.csiro.au/observing/documentation/user_guide/pks_ug_3.html}).
\item $^\mathrm{(a)}$ in the incoherently and coherently beamformed data, respectively.
\item $^\mathrm{(b)}$ in the coherently beamformed data.

\end{tablenotes}

\end{table*}

\subsection{Localisation}
\label{sec:sub:localisation}
The pulsar was observed on four consecutive days in May 2018 with GBT, spread out evenly over the sdB orbit to cover all orbital phases with the aim to detect the acceleration of the pulsar in its expected orbit
around the sdB and to start a timing solution to localise it precisely. Interestingly, the pulsar was detected in only one of these four observations.
In the detection data, there was, again, no hint of acceleration.
By itself this does not \emph{rule out} the association:
  The binary could also be more face-on than assumed from the
  optical radial velocity curves,
  or the difference in mass between the two binary companions may have been larger than assumed.
  But the opposite, a detection of the acceleration, could have immediately associated the pulsar firmly with the sdB star.
Due to the high fraction of non-detections, there 
were not enough data points to localise the pulsar through timing, either.
We ruled out a number of reasons for the non-detections:
  the RFI situation between the four were similar and the test pulsar was detected equally well in all, indicating our sensitivity in all four was the same (Table~\ref{tab:targets});
  the orbital phases between the four were significantly different, ruling out that eclipses play a major role; and we searched in period, period
  derivative, and dispersion measure. After eliminating these causes, the most probable remaining reason for the non-detections
  was intrinsic nulling or moding behaviour in the pulsar.

We then proceeded to observe the pulsar with uGMRT using several observing modes simultaneously. Beamformed data were recorded using both an incoherent addition of typically 16 dishes, as well as a 
coherent addition of the central 12 antennas. The incoherent mode retains the half-power beam width of a single uGMRT dish, ${\sim} 70\arcmin$, which covers the full field as observed with GBT. The coherent
mode has a half-power beam width of ${\sim}5\arcmin$, but is a factor three more sensitive than the incoherent mode. Since the sdB system was used as the pointing centre, the pulsar would be in the centre of the beam if it were part of the sdB system, and hence
have a higher $S/N$ in the coherent data than in the incoherent data. The pulsar was indeed detected, however the $S/N$
was three times higher in the \emph{incoherently} beam-formed data. The test pulsar did have a higher $S/N$ in the
 coherent data, so the system performed as expected. Hence, we conclude that the pulsar is not associated with the sdB binary system. 
 
\begin{figure*}
    \centering
    \includegraphics[width=.9\textwidth]{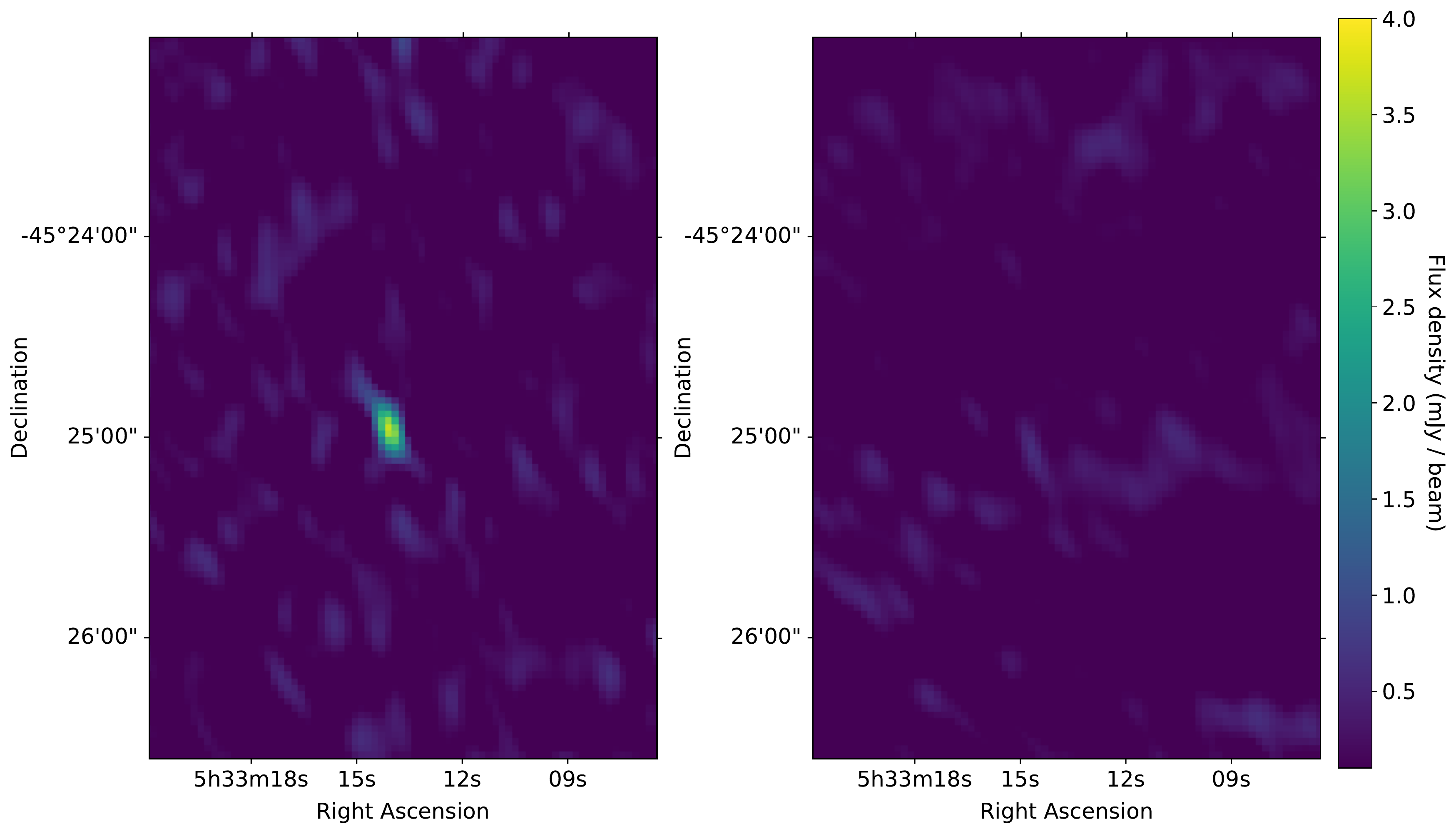}
    \caption{uGMRT images from observing sessions with detection on 20161106 (left) and non-detection on 20181103 (right) of the pulsar in beamformed mode. The source shown here is the only source of which the flux density in the image correlates with the flux density of the pulsar in beamformed mode, hence we assume it is in fact the pulsar.}
    \label{fig:HE0532image}
\end{figure*}

In addition to beamformed data, interferometric data were recorded. Using the hypothesised nulling behaviour of the pulsar to our advantage, we aimed to image the field of both an observation with a detection and non-detection of the pulsar in beamformed mode. Any source in the image that shows the same on/off behaviour and has a flux density consistent with the flux density measured in beamformed data, might be the pulsar. The image created from the 20161106 uGMRT observation contained the pulsar in its on state (Fig.~\ref{fig:HE0532image}).

An off-state image was made from the data taken on 20181103. We identified one source, at RA = 05:33:14, Dec = $-$45:24:50, that was only present in the on-state image. The detection and non-detection images are shown in Fig.~\ref{fig:HE0532image}. This refined position was used for the last two uGMRT follow-up observations on 20190201 and 20190301.

In the 20190201 observation, the pulsar signal was clearly detected with an integrated $S/N$ of 50 and corresponding average flux density of $0.58(10)\,\mathrm{mJy}$, which was the most significant detection thus far. In addition, over 100 single pulses were detected. We are thus confident that the source
identified in the image is indeed the pulsar. Due to issues with processing of the interferometric data, the images created from the 20190201 and 20190301 observations did not have enough
sensitivity to be able to identify the pulsar. The pulsar position is ${\sim}20\arcmin$ from the sdB position and localised to ${<}1\arcmin$, hence we conclude the pulsar and sdB are not associated.

The pulsar was also observed with Parkes in April of 2016 for 6.7 hrs but no periodic signal nor single pulses were found. As the half-power beam radius is ${<}8\arcmin$ for Parkes' H-OH receiver at $1.4\,$GHz, the observation had little sensitivity towards the pulsar position, giving an upper limit of $45(9)\,\mathrm{mJy}$\footnote{Using $G = 0.83\,$K/Jy and $T_\mathrm{rec} = 25\,$K from the Parkes user guide (\url{https://www.parkes.atnf.csiro.au/observing/documentation/user_guide/pks_ug_3.html}}. Given also the unknown pulsar spectral index, the non-detection is not surprising. We therefore did not consider this observation further.

\subsection{Giant pulse emission}
\label{sec:sub:j0533_gp}
There are several definitions of giant pulses, but a broadly accepted one is any pulse that has a period-averaged flux density that is at least ten times higher than the mean flux density of the periodic signal \citep{jr04a,Cairns04,ksv10,SV12,Maan19b}. They are also narrower than the integrated profile and sometimes occur in a very narrow phase window \citep{kni06}.
To classify single pulses from \J, we considered the 20190201 uGMRT observation, which is the only observation with a periodic detection and the source in the centre of the beam.

The $S/N$ of individual pulses reported by the {\sc single\_pulse\_search.py} 
tool from {\sc PRESTO}
already correspond to a downsampling in time which maximises their S/N.
This is equivalent to defining the width as the width of a top-hat with the same peak and integrated $S/N$ as the observed pulse. 
We use this downsampling factor as an approximation for the
actual pulse-width. The sky background temperature towards the
source is estimated to be 17\,K \citep{hssw82}. We have assumed
$T_\mathrm{rec}$ to be 125\,K, implying a total $T_\mathrm{sys}$
of 142\,K (the observatory specifies a $T_\mathrm{sys}$ range of 100$-$165).
We used these parameters in the modified
radiometer equation to compute the peak flux density \citep{CM03,MA14} of
individual pulses.

\begin{figure}
    \centering
    \includegraphics[width=\columnwidth]{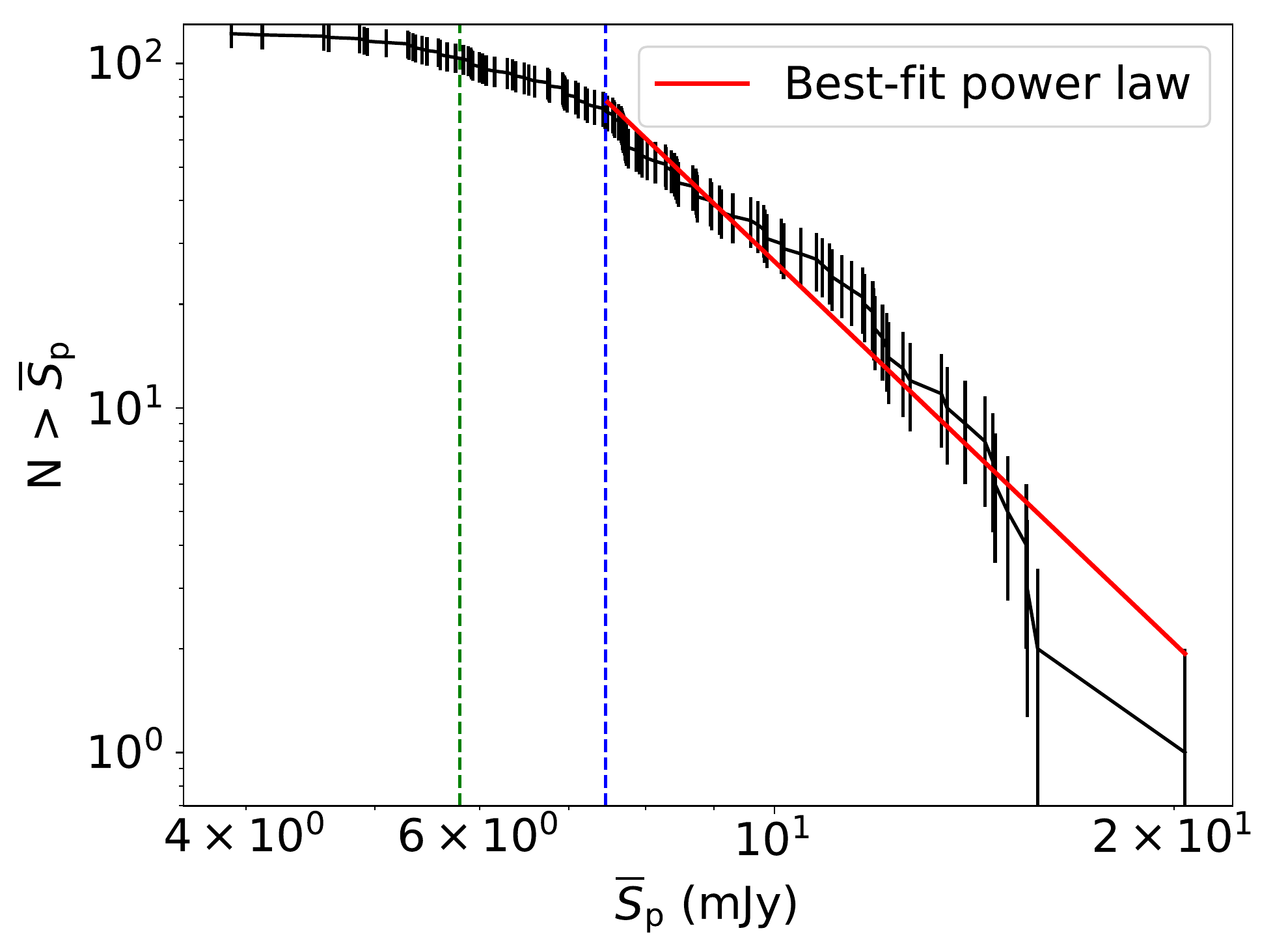}
    \caption{Cumulative distribution of period-averaged flux density of detected single pulses in log-log space. The vertical dashed lines indicate the giant pulse threshold of 10 times the mean flux density ($5.8\,$mJy, green) and completeness threshold ($7.5\,$mJy, blue). The best-fit power law for all pulses above the completeness threshold (red) has a slope of $-3.68(1)$.} 
    \label{fig:j0533_gp_cdf}
\end{figure}

To compare the derived peak flux densities to the flux density of the periodic signal we define the period-averaged flux density of a single pulse as $\overline{S}_p = S_\mathrm{p} \times W/P$, where $S_\mathrm{p}$ is the peak flux density, $W$ is the width of the single pulse and $P$ is the period of the pulsar. $\overline{S}_p$ incorporates any differences in width between the periodic profile and single pulses, which makes it the appropriate parameter for comparison between single pulses and the periodic signal.
The cumulative distribution function (CDF) of detected single pulses is shown in Fig.~\ref{fig:j0533_gp_cdf}. The giant pulse threshold of ten times the mean flux density of $0.58\,$mJy (cf. Table~\ref{tab:he0532_followup}) is shown as dashed green line. $92\%$ of the detected single pulses are above this threshold. Hence they are consistent with being giant pulses. 

Even assuming we are complete down to a $S/N$ of 8, the completeness in $S_\mathrm{p}$ depends on the pulse width. If the widest observed pulse were detected at $S/N = 8$, it would have a peak flux density of $7.5\,$mJy. We take this value as our completeness threshold. The slope of the best-fit power law to the pulses above the completeness threshold is $-3.68(1)$.

\begin{figure}
    \centering
    \includegraphics[width=\columnwidth]{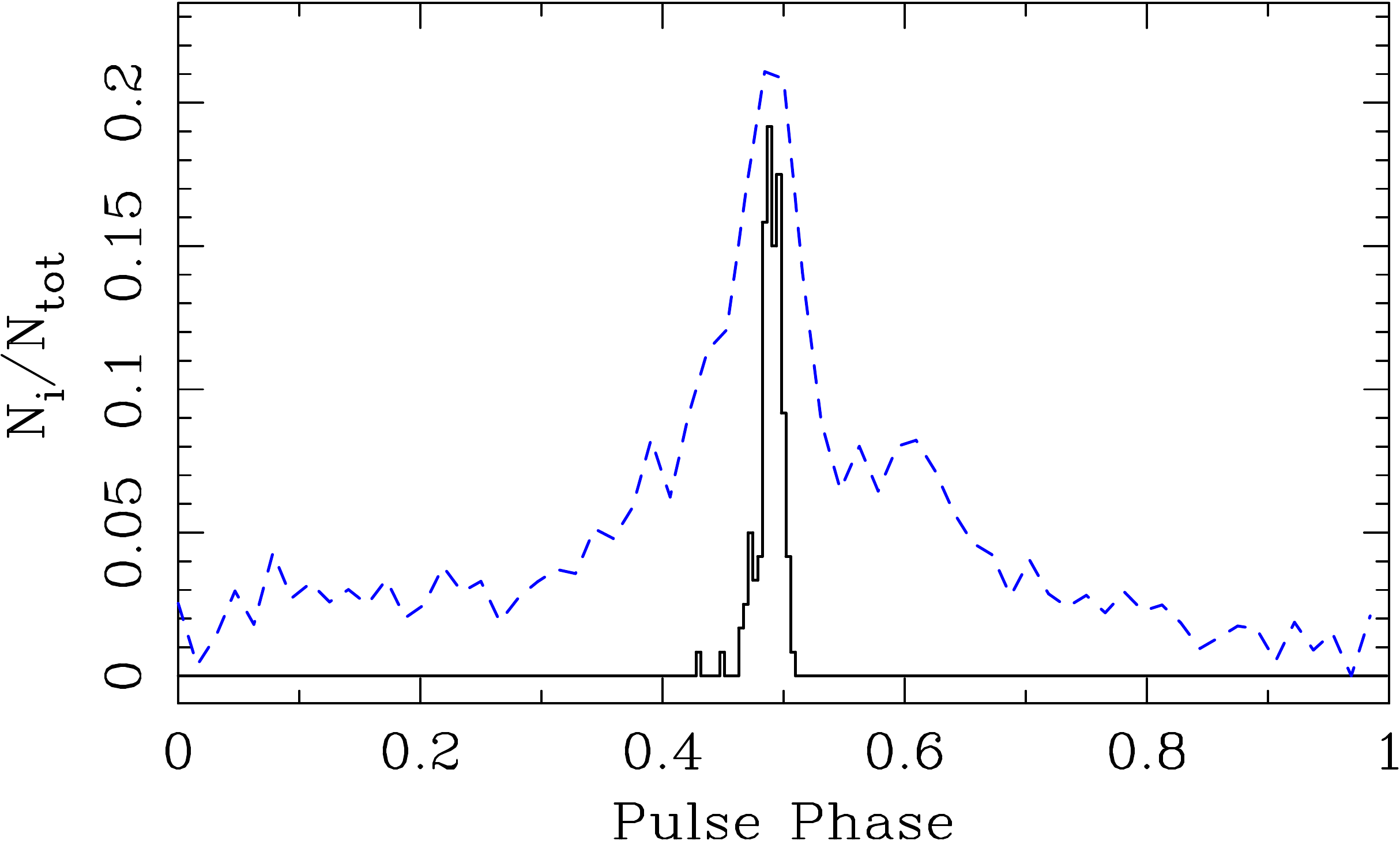}
    \caption{Distribution of rotational phase of single pulses (black) overlaid on the integrated pulse profile (blue). The giant pulse occur in a very narrow phase window around the peak of the integrated profile. They do not occur in the trailing edge component of the integrated profile.}
    \label{fig:j0533_gp_phases}
\end{figure}

Using the barycentric period measured in the 20190201 observation with {\sc prepfold} from {\sc PRESTO}, the barycentric arrival time of each of the 120 single pulses was converted to a rotational phase. A histogram of 
the resulting phases is shown in Fig.~\ref{fig:j0533_gp_phases}, with the integrated profile shown for reference. All pulses occur within a phase window of 0.04, where the peak matches that of the integrated profile. They do not occur in the trailing component of the integrated profile. The giant pulses have widths between $2.5$ and $10.0$\,ms, which is 10-30\% of the width of the integrated profile ($24 $\,ms), where the width is defined as the width of a top-hat with the same peak value and integrated flux density as the observed pulse. 
These widths are a similar fraction of the mean pulse as the giant pulses observed in PSR~B0950+08 \citet{tsa+15}. Together with the narrow phase window centred on one component of the integrated profile, this supports that the single pulses are indeed giant pulses \citep{kni06}.

\subsection{Nulling and mode changing}
The initial seemingly erratic series of detections and non-detections (see Table~\ref{tab:he0532_followup})
were reminiscent of the struggle to confirm and study mode-changing, nulling, and intermittent pulsars as
PSRs B1931+24 \citep{klo+06}, B0826$-$34 \citep{lt12}, and J1929+135 \citep{2017ApJ...834...72L}.

\begin{figure*}
    \centering
    \includegraphics[width=\textwidth]{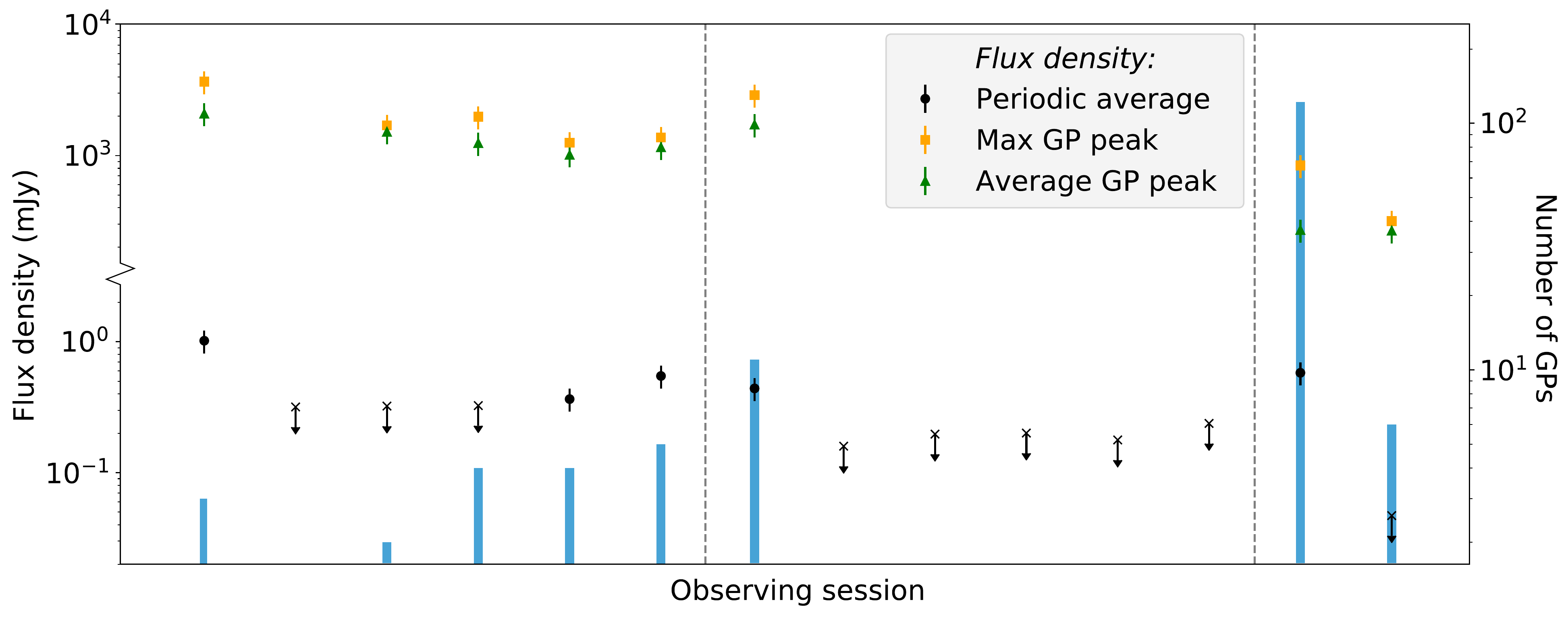}
    \vspace{-5mm}
    \caption{Overview of periodic and single pulse behaviour in PSR \J. 
    The periodic average flux density is shown as black dots, with
    upper limits shown as crosses. The average and brightest giant pulse flux density are shown in green and orange, respectively. Error bars indicate $1\sigma$ errors. A histogram of the number of detected giant pulses in each session is shown in blue. The vertical dashed lines indicate changes to a different observing setup: The first six sessions were with GBT, sessions seven through twelve are based on uGMRT in incoherent mode, all with the pulsar not at boresight. The last two sessions were with uGMRT in coherent mode, with the pulsar at boresight.}
    \label{fig:j0533_possible_modes}
\end{figure*}

The disappearance of the source near the end of the discovery observation (see Fig.~\ref{fig:J0533_discovery}) suggests that its flux density decreased, but it was observed very close to the horizon and might have set instead. Part of the variation in observed flux density in different observations is due to the initial positional uncertainty, and mis-pointing. 
But sets of detections using similar telescope setups can be compared among themselves, to analyse if intrinsic mode
changing is also at play. 
In Fig.~\ref{fig:j0533_possible_modes} we visualise the period and single-pulse detections.
Sets demarcated by dashed lines were observed with the same setup and can be meaningfully compared.

We see that the periodic average flux density for observations \emph{with} detections
is only a factor of a few above our upper limits for non-detections.
The fact that the initial detection is brighter than average can be explained by a discovery bias.
Only in the last epoch, 
using the coherent uGMRT at boresight, there is a factor of 10 difference between the periodic average flux density and non-detection upper limit. 
In known nulling and mode-changing pulsars such as
B0809+74 \citep{lkr+02} and
B0826$-$34,
the flux density at the source changes by a factor of order 50 \citep{2005MNRAS.356...59E}.

In the first set of observations, the giant-pulse occurrence rate and peak flux density do not appear to correlate with whether
periodic emission is detected. In the second set, they do.
There, each observation either delivered the detection of both periodic and giant-pulse emission, or of neither.

Overall, the difference between our detections and upper limits
  does not reach the brightness difference of several orders of magnitude generally seen in nulls or between modes.
We thus conclude our data \emph{suggest} nulling,
but are not constraining enough to \emph{prove} nulling or mode changing.

\subsection{Timing}
\label{sec:sub:timing}
In order to characterise the pulsar parameters, we aimed to create a coherent timing solution. For several observations, only single pulses are detected. As the single pulses occur in a very narrow phase window around the peak of the integrated pulse, both the single pulse and periodic arrival times can be used to form a 
timing solution.
For both the periodic profile and single pulses a template profile was created using 
{\sc dspsr} and {\sc psrchive}, based on the highest $S/N$ detections. Times-of-arrival (TOAs) were then extracted from each single pulse, as well as from each periodic detection. For both single pulse and the periodic signal we used a pulse profile template with the same phase and shape, but with a different width determined from the highest S/N data.
For observations where the periodic $S/N$ was high enough, the observation was split into chunks of at least $S/N$ 8 each and TOAs were extracted for each chunk.

We then proceeded timing with {\sc TEMPO2}. 
%Initially, the timing was based on the position of the sdB star. A coherent solution could not be found. We tried excluding the 2011 data as there is a large gap without data between that data set and the next, but to no avail. Howver,
When the position was updated to the variable source discovered in the imaging data, it was possible to find a coherent solution for the 2016 -- 2019 data. The 2011 points then also fit the solution well, so they were included in the analysis. Then, the DM was fit by splitting the highest $S/N$ periodic detection into 32 frequency chunks and fitting with {\sc TEMPO2}. Finally, the position was then allowed to vary as well. The final derived position is consistent with that measured from the imaging technique.
The fit parameters are shown in Table~\ref{tab:timing_sol}. 

The DM suggests a distance of $0.7\,$kpc using the NE2001 electron model \citep{cl02} and $1.3\,$kpc using YMW16 \citep{ymw16}
The obtained period of $157.28\,$ms and $\dot{P}$ of $2.8\times10^{-16}$ suggest a characteristic age (defined as $P / 2\dot{P}$) of ${\sim} 10\,$Myr and surface magnetic field (defined as $10^{12}\sqrt{P \dot{P}}\,\mathrm{G}$, with $P$ in seconds) of ${\sim} 2\times10^{11}\,$G. The pulsar is thus a bit older than one might expect given its period, but it has a relatively low magnetic field. These parameters are similar to PSR~B0950+08, which has a period of $253\,$ms, and period derivative of $2.3\times10^{-16}$ \citep{hlk+04}. 

\begin{figure}
    \centering
    \includegraphics[width=\columnwidth]{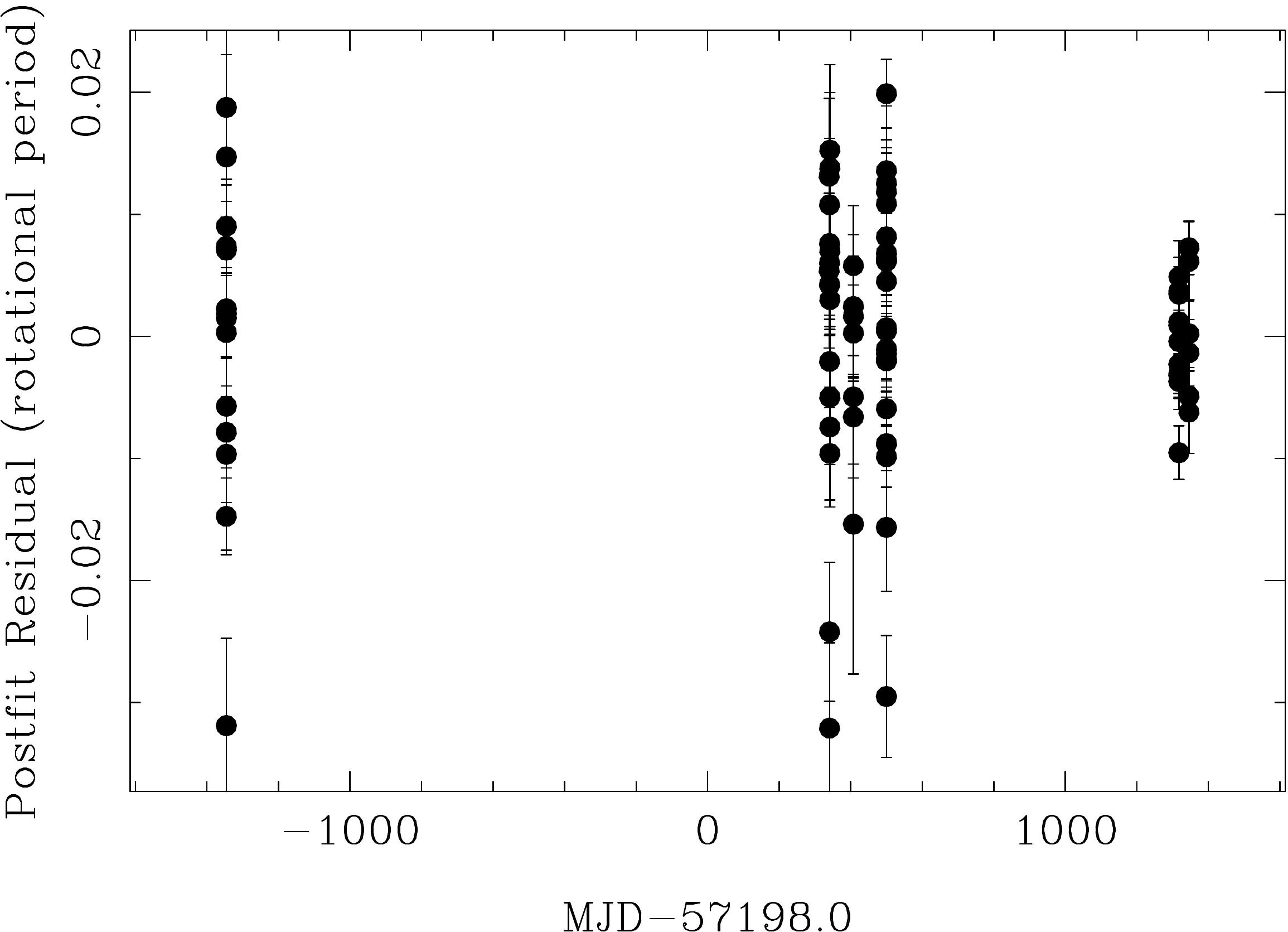}
    \caption{Timing residuals of PSR \J. We obtained 82 (25 periodic and 57 single-pulse) arrival times spread over ${\sim}7$ years. The mean residual is ${<}1\%$ of the pulse period. See also Table~\ref{tab:timing_sol}}
    \label{fig:timing_res}
\end{figure}

\begin{table}
    \caption{Parameters of the best-fit timing solution determined with {\sc TEMPO2}.}
    \label{tab:timing_sol}
    \centering
    \begin{tabular}{ll} \hline
         \emph{Fit and data-set} & \\
         Pulsar name & PSR J0533$-$4524\\ 
         MJD range &  55852 -- 58544\\ 
         Weighted RMS timing residual$^\mathrm{(a)}$ ($\mathrm{\mu s}$) & 1105.453\\ 
         $\chi^2_\mathrm{red}$ & 6.0 \\\hline
         \emph{Set quantities}  & \\ 
         Period epoch (MJD) & 57605.55728\\  \hline
         \emph{Measured quantities} & \\
         DM ($\pccm$) & 18.93(2) \\
         RA & 05:33:13.89(4)\\ 
         Dec & -45:24:50.2(2)\\
         $P$ (s) & 0.157284525096(2) \\
         $\dot{P}$ (s/s) & 2.8024(2)$\,\times\,10^{-16}$ \\ \hline
         \emph{Derived quantities} & \\
         Characteristic age (Myr) & 8.8925(4) \\
         B$_\mathrm{surf}$ (G) & 2.1245(1) $\,\times\,10^{11}$ \\ \hline 
    \end{tabular}
\begin{tablenotes}
\item $^\mathrm{(a)}$ Defined as the root-mean-square deviation from zero of each residual, weighted by its uncertainty.
\end{tablenotes}
\end{table}

\section{Discussion}\label{sec:discussion}
In Sect.~\ref{sec:resultssdb} we have shown that given the derived beaming fraction of known MSPs,
and their place in the luminosity distribution,
at most three out of six systems can be expected to be millisecond radio pulsars.
Furthermore, all three would be beamed away from us or too dim to be detected.
It might also be possible that the pulsars are only mildly recycled, and their beams have a significant chance
to miss Earth.
In both cases we cannot exclude that they are all neutron stars, just not detectable in radio. 

To establish whether it is reasonable to assume all six systems host a
neutron star, we consider two aspects: the neutron star birth rate and their behaviour in binary systems.

Supernova modelling already indicates a lower rate of neutron-star formation than
  appears to be required to produce the
number of pulsars observed \citep[cf.][]{kk08}.
Is this problem  twice as bad, if half of neutron stars are not detectable as radio pulsars, as our observations
seem to suggest? Not directly.
The neutron-star  birth rate for successful modelling of the Galactic population, in such population synthesis as
\citet{fk06} and \citet{ls10}, is only that of regular, non-recycled pulsars.
These first shine during their regular lives, to then possibly be reborn as MSPs.
Systems that will later evolve into systems like our six  may  currently be visible as regular pulsars,
where they are properly counted toward the neutron-star birth-rate problem. 
The closest such system, i.e., a currently observed pulsar that may later evolve into an sdB-PSR binary, 
is PSR J0045$-$7319 \citep{kjb+94}, and there are two similar but more massive known binaries.

The Small Magellanic Cloud pulsar PSR J0045$-$7319
has a companion of type B1\,V, of 8-10\,M$_\odot$ \citep{kjb+94}.
The pulsar there is not yet recycled and the orbit is still 51\,days.
For a B-type companion of such ${>}5$\,M$_{\odot}$ mass \citet{2018A&A...618A..14W} predict common envelope evolution, with
short (${\sim}$hour) orbits, similar to the six candidate systems investigated in this work (Table~\ref{tab:targets}).

The binary companion to PSR B1259$-$63 \citep{jml+92} is now thought to have a mass of 15-31 M$_\odot$
\citep{2018MNRAS.479.4849M}, which may be too large to become an sdB star. While it was thought to be a Be star for the first few years
after its discovery, it is currently classified as Oe star.
Similarly, the latest timing on  PSR J1740$-$3052 indicates its companion has a mass of 16-26 M$_\odot$
\citep{2012MNRAS.425.2378M}, and the most likely optical counterpart is a main sequence star of late O or early B type
\citep{bbn+11}.

Together, these three observed systems qualitatively suggest our non-detections do not immediately create a birth-rate
problem. Is the recycling process perhaps not reliable?

In the previous discussion we have estimated the beaming fraction $f_b$,
the odds that the beam of an active MSP sweeps across Earth.
But what is the fraction  $f_r$ of systems that is successfully recycled?
As the beaming fraction allows for 3 of 6 systems to go unseen,
 we conclude  an additional factor  $f_r{<}0.5$ must be needed to explain our non-detections.
Such a fraction significantly smaller than 1 is in line with
the number of radio non-detections in other systems where neutron stars could viably be present as pulsars.
These targeted searches included binaries such as low-mass white dwarfs \citep{lfm+07,2009ApJ...697..283A},
OB runaway stars \citep{1996ApJ...461..357S}, and soft X-ray transients \citep{2017ApJ...840....9M}.
Even the radio detection of PSR~J1417$-$4402 by \citet{2016ApJ...820....6C}
appears to have occurred independently from the optical identification of the binary 1FGL J1417.7$-$4407 \citep{2015ApJ...804L..12S}.

It remains possible that all observed sdB binaries in fact host a neutron star, but all of them would either be beamed away 
from Earth, too dim, or not recycled. Could it be possible that some systems actually host a white dwarf instead of a 
neutron star? This would mean that their masses must be below $1.4\Msun$. The masses are determined under the assumption of co-rotation, so the orbital period of the system is assumed to be equal to the rotation period of the sdB. This allows for determination of the inclination angle and hence the mass ratio of the two components. As the range of masses allowed for sdB stars is quite small, this gives the mass of the secondary to reasonable precision. If some of the suspected neutron stars are actually white dwarfs, their masses must have been overestimated. Getting to the right mass range would require either the sdB mass to be much smaller, which seems nonphysical, or the derived inclination angle to be too high, which could happen if the assumption of co-rotation breaks down. If these systems are actually more edge on, the predicted masses would be lower. This would also solve the inclination problem posed by \citet{geier+2010}. We note that for a random distribution of inclination angles, the most probable value is $52\degr$, which puts the predicted secondary masses in the 0.9-1.0$\,\Msun$ range. It therefore seems likely that several of the observed sdB system actually host white dwarfs if the assumption of co-rotation does not hold. Only PG\,1232$-$136 still has a predicted mass of ${>}1.4\,\Msun$ and remains a viable system to host either a neutron star or black hole. 

X-ray emission may be expected from sdB-NS systems due to thermal emission of the neutron star or accretion of the sdB wind \citep{mereghetti2011}. However, targeted searches for X-ray emission from our six targets have not yielded any detections \citep{mereghetti2011,mpe+14}. This further suggests the absence of neutron stars although it might also be explained by mass-loss rates that are lower than predicted by theoretical models. These non-detections also confirm that there is no significant accretion due to Roche-Lobe overflow, which is expected given that sdBs are much smaller than their Roche Lobe.

\subsection{PSR J0533$-$4524}

\subsubsection{Is PSR J0533$-$4524 an RRAT}
We observe pulsar \J often emits strong individual pulses.
Should it then be classified as a rotating radio transient (RRAT)?
According to the definition proposed in \citet{mll+06}, one of the characteristics of an RRAT is
that its period is determined from the single pulses, and
cannot be derived from periodic emission.
Initial observations fit this definition. But, as
 we were able to measure the period from the Fourier search on the 20111018 observation,
\J is ultimately not an RRAT.

\subsubsection{Giant pulse emission revisited}
While the single pulses detected from \J are giant pulses according to the typical definition, we consider they might
be the bright end of a single underlying single-pulse distribution, as was determined for PSR B0950+08 \citep{tsa+16}. While
for B0950+08, the underlying distribution is assumed to be Gaussian, \citet{kjv02} showed that several pulsars have a
log-normal pulse brightness distribution. 

Assuming \J has a log-normal distribution of single pulses, we can predict the slope of the observed CDF of single pulses without fully knowing the underlying single-pulse distribution. The fraction of detectable single pulses, $f_\mathrm{sp}$, is equal to the chance of detecting a pulse that is more than $n \sigma$ brighter than the mean pulse ($\mu$) for some unknown $n$, and is given by the complement of the CDF of the log-normal distribution,
\begin{equation}
    \label{eq:erf}
    f_\mathrm{sp} = \frac{1}{2} \erfc{(\frac{n}{\sqrt{2}})}\,,
\end{equation}
where $\erfc$ is the complementary error function.

The slope of the CDF of detected single pulses is then given by the derivative of Eq.~\ref{eq:erf}. Rewriting in terms of $f_\mathrm{sp}$ gives 
\begin{equation}
    \label{eq:slope}
    \frac{\partial \log f_\mathrm{sp}}{\partial \log n} = \frac{-\erfcinv{(2f_\mathrm{sp})}}{f_\mathrm{sp}\sqrt{\pi}} \,\exp{\left(-\erfcinv{(2f_\mathrm{sp})}^2\right)}\,,
\end{equation}
where $\erfcinv$ is the inverse of the complementary error function. The slope predicted by this equation is equal to the slope of the observed CDF \emph{if} the distribution of the parameter that is chosen to create the CDF has a mean of zero. Evidently, this is not the case if the chosen parameter is the period-averaged flux density of the single pulses, $\overline{S}_\mathrm{p}$. Instead, we choose $\log_{10}(\overline{S}_\mathrm{p} / S_\mathrm{p, mean})$, where $S_\mathrm{p, mean}$ is the mean flux density of the periodic profile ($0.58$\,mJy, see Table~\ref{tab:he0532_followup}). The mean of the distribution then is equal to zero if there is indeed one underlying single-pulse distribution. 

\begin{figure}
    \centering
    \includegraphics[width=\columnwidth]{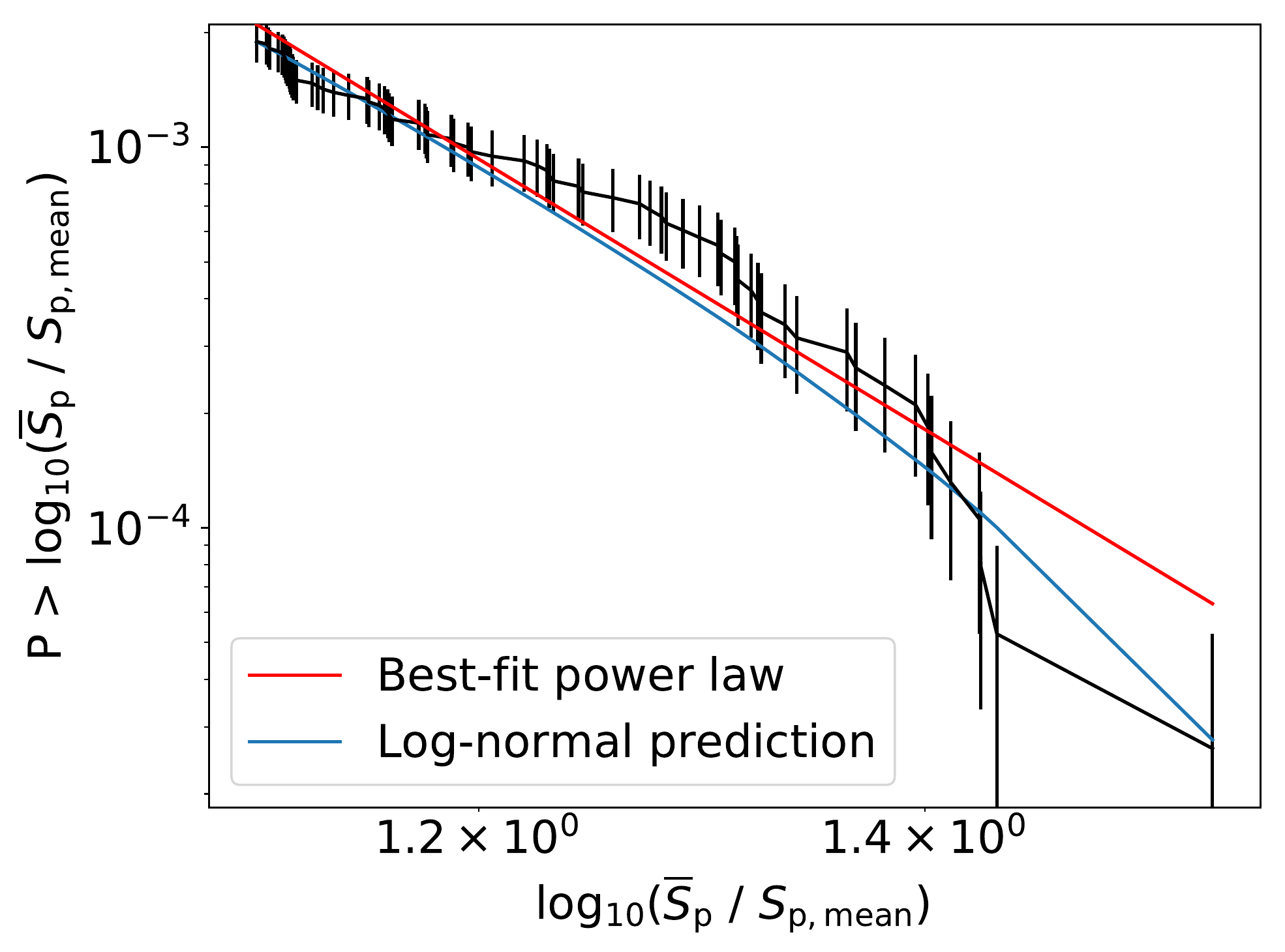}
    \caption{Cumulative distribution of $\log_{10}(\overline{S}_\mathrm{p} / \overline{S}_\mathrm{p, mean})$ of detected single pulses. The best-fit power law is shown. The slope is $-10.6(4)$ in log-log space, which is consistent with the value predicted for the tail of a log-normal distribution of single pulses ($-11$).}
    \label{fig:j0533_gp_cdf_analysis}
\end{figure}

The observed distribution of $\log_{10}(\overline{S}_\mathrm{p} / S_\mathrm{p, mean})$ is shown in Fig.~\ref{fig:j0533_gp_cdf_analysis}. $1\sigma$ error bars are shown assuming Poissonian errors. There are 72 single pulse above the completeness threshold defined in Sect.~\ref{sec:sub:j0533_gp}. In total, the pulsar has $3.8\times10^{4}$ turns in the 1.7-hr part of the observation where it was visible, implying $f_\mathrm{sp} = 1.9\times10^{-3}$, which corresponds to detecting all single pulses that are at least $2.9\sigma$ brighter than the mean pulse. This also implies that the standard deviation of the underlying distribution is ${\sim}0.38$ in units of $\log_{10}(\overline{S}_\mathrm{p} / S_\mathrm{p, mean})$.

Equation~\ref{eq:slope} then predicts a slope of $-9.2$ for the CDF of detected single pulses at the completeness threshold. Extrapolating from this point, the predicted CDF is shown in blue. It has a mean slope of $-11$. The best-fit power law is shown in red and has a slope of $-10.6(4)$. The observed distribution is consistent with being the bright-end tail of a log-normal distribution of single pulses.

It is thus not straightforward to classify single pulses. In some cases, giant pulses may simply be the bright end of the distribution of normal pulses. However, their width and phase are different from those of the average pulse and when averaged together they do not recover the periodic integrated profile. More research into this subject is needed to determine whether this means the \J giant pulses are actually from a different distribution than the normal pulses, or whether it implies a correlation between these parameters and pulse brightness. A correlation between pulse width and brightness is seen in for example the Crab giant pulses \citep{ksv10}, where narrower pulses are typically brighter. 

If giant pulses are simply the tail of the normal single-pulse distribution, then why are they not detected in all
pulsars given enough observation time? This may be due to differences in the width of the normal pulse distribution. For
\J, a pulse that is $3\sigma$ above the mean is roughly ten times brighter than the mean pulse, and hence classified as a
giant pulse. If, however, the single-pulse distribution were narrower, a pulse with ten times the mean flux density would be much more
rare. For example, the single-pulse distribution of a standard pulsar such as PSR B0818$-$13 in its on state \citep{jl04} has virtually no pulses that are more than twice as bright as the mean. Perhaps several giant-pulse emitting pulsars are classified as such because they have a relatively broad single-pulse distribution, such that it is feasible to detect pulses over ten times the mean within a typical observation length.

\section{Conclusions}\label{sec:conclusions}
We searched for radio pulsations from six sdB binary systems that are likely to host a neutron star based on sdB orbital parameters derived from optical observations.
No pulsars were detected towards sdB systems down to an average flux density limit of $0.13\,$mJy at $350\,$MHz. For the systems also presented in \citet{cls11}, the upper limits are a factor 2-3 deeper. The non-detection of any pulsar towards the sdB binary systems could be explained by a combination of the putative MSP beaming fraction, luminosity, and a recycling fraction $f_r {<} 0.5$. If some of the sdBs systems do host a pulsar, the most likely reason for non-detection is that their beams do no sweep across the Earth. Therefore it is unlikely that deeper searches, either through longer observations or through the use of more sensitive telescopes, such as Arecibo and FAST, will be fruitful. However, it is possible that there might be a pulsar with extremely faint emission, which may be detectable with the aforementioned instruments.
It is also possible that the assumption of co-rotation of the sdB in its orbit does not hold, in which case the masses of the sdB companions are likely over-predicted. Then, several systems could host a white dwarf instead of a neutron star.

We discovered PSR \J, a giant-pulse emitting pulsar. Through simultaneous beamformed and interferometric observations with uGMRT, the pulsar was localised and shown to be a serendipitous discovery, not associated with the sdB system that was the original target. We detected over 100 giant pulses from this pulsar. Their distribution is compatible with the tail of a log-normal distribution with the same mean as the average single pulse, showing that we may be seeing the bright end of the normal single pulses. However, the giant pulses are narrower than the integrated pulse and restricted to a very narrow phase window, unlike what is expected from the average single pulses.

\section*{Acknowledgements}
We thank J.~Lazio for providing the data on HE\,0929$-$0424 and PG\,1232$-$136 and E.~Petroff for observing HE\,0532$-$4503 with
Parkes. Additionally, we thank the anonymous referee for providing useful comments that helped improve the manuscript.
LCO, JvL, and YM acknowledge
funding from the European Research Council under the European Union's Seventh Framework Programme (FP/2007-2013)/ERC Grant Agreement No. 617199. 
The Green Bank Observatory is a facility of the National Science Foundation operated under cooperative agreement by Associated Universities, Inc. The Westerbork Synthesis Radio Telescope is operated by ASTRON (The Netherlands Institute for Radio Astronomy) with support from the Netherlands Foundation for Scientific Research (NWO).
GMRT is run by the National Centre for Radio Astrophysics of the Tata Institute of Fundamental Research.

\bibliographystyle{mnras}
\bibliography{journals_apj,sdB,psrrefs,modjoeri,modrefs,refthijs}

\end{document}